\documentclass[sigconf,nonacm]{acmart}

\setcopyright{none}
\renewcommand\footnotetextcopyrightpermission[1]{}
\usepackage{makecell}
\usepackage{xcolor}
\usepackage{listings}
\usepackage{xspace}
\usepackage{todonotes}

\definecolor{codegreen}{rgb}{0,0.6,0}
\definecolor{codegray}{rgb}{0.5,0.5,0.5}
\definecolor{codepurple}{rgb}{0.58,0,0.82}
\definecolor{backcolour}{rgb}{0.95,0.95,0.92}

\lstdefinestyle{pystyle}{
    language=Python,
    frame=single,
    commentstyle=\color{codegreen},
    keywordstyle=\color{magenta},
    numberstyle=\tiny\color{codegray},
    stringstyle=\color{codepurple},
    basicstyle=\ttfamily\footnotesize,
    breakatwhitespace=false,
    breaklines=true,
    captionpos=b,
    keepspaces=true,
    numbers=left,
    numbersep=5pt,
    showspaces=false,
    showstringspaces=false,
    showtabs=false,
    tabsize=2
}

\AtBeginDocument{%
  }

\newcommand{\linuxdata}{\textsf{\textit{TuxKConfig \xspace}}}

\acmConference[EASE 2025]{The 29th International Conference on Evaluation and Assessment in Software Engineering}{17–20 June, 2025}{Istanbul, Türkiye}

\begin{document}
\setcopyright{none}


\title[Linux Kernel Configurations at Scale: A Dataset for Performance and Evolution Analysis]{Linux Kernel Configurations at Scale:\\ A Dataset for Performance and Evolution Analysis}

\author{Heraldo Borges}
\affiliation{%
 \institution{Univ Rennes}
 \institution{Inria, CNRS, IRISA}
  \city{Rennes}
  \country{France}}
  \email{heraldo.pimenta-borges-filho@irisa.fr}

\author{Juliana Alves Pereira}
\affiliation{%
 \institution{Pontifical Catholic University of Rio de Janeiro}
   \city{Rio de Janeiro}
  \country{Brazil}}
  \email{juliana@inf.puc-rio.br}

  \author{Djamel Eddine Khelladi}
\affiliation{%
 \institution{Univ Rennes}
 \institution{Inria, CNRS, IRISA}
  \city{Rennes}
  \country{France}}
  \email{djamel-eddine.khelladi@irisa.fr}

\author{Mathieu Acher}
\affiliation{%
 \institution{Univ Rennes}
 \institution{Inria, CNRS, IRISA}
  \city{Rennes}
  \country{France}}
  \email{mathieu.acher@irisa.fr}

\renewcommand{\shortauthors}{Heraldo Borges, Juliana Alves Pereira, Djamel Eddine Khelladi and Mathieu Acher}

\begin{abstract}
  Configuring the Linux kernel to meet specific requirements, such as binary size, is highly challenging due to its immense complexity—with over 15,000 interdependent options evolving rapidly between versions. Although several studies have explored sampling strategies and machine learning methods to understand and predict the impact of configuration options, a comprehensive and large-scale dataset that encompasses multiple kernel versions, along with detailed quantitative measurements, remains lacking in the literature. To bridge this gap, we introduce \linuxdata, an accessible collection of kernel configurations that spans multiple kernel releases (including versions from 4.13 to 5.8). This dataset, gathered through automated tools and builds, comprises more than 240,000 kernel configurations systematically labeled with compilation outcomes and binary sizes. By providing detailed records of configuration evolution and the intricate interplay among kernel options, our dataset supports innovative research in feature subset selection, machine learning-based prediction models, and transfer learning between kernel versions. Throughout the paper, we describe how the dataset has been made easily accessible via OpenML and demonstrate how it can be used with just a few lines of Python code to assess AI-based techniques (e.g., supervised machine learning).
  We expect this dataset to enhance reproducibility and drive new insights into configuration-space analysis at a scale that presents a unique opportunity and inherent challenges, advancing our understanding of the Linux kernel's configurability and evolution.

\end{abstract}



\maketitle

\fancyhead{}
\fancyhead[LE]{\shorttitle}
\fancyhead[RO]{\shortauthors}

\section{Introduction}


Ensuring that software systems meet specific functional and non-functional requirements—such as binary size—is critical in software engineering. This challenge is particularly pronounced in highly configurable systems like the Linux kernel, which includes over 15,000 interdependent options evolving rapidly across versions, with use-cases spanning IoT devices and supercomputers~\cite{Abal2014LinuxVariabilityBugs, martin2021a, acher2022}. For instance, in embedded systems such as IoT devices, reducing the kernel's binary size is crucial for operation on hardware with limited memory, while in high-performance servers, compilation time can be a critical bottleneck in development cycles.
Beyond Linux, practitioners often face difficulties selecting optimal configurations due to the combinatorial explosion of possible configurations and subtle, unpredictable interactions among these options~\cite{pereira2020a, alvespereira:hal-02148791, sarkar2015, jamshidi2017, Abal2018VariabilityBugs, nair2020, quinton2020evolution,beetle, DBLP:conf/wosp/ValovPGFC17, nair2017, FSE2017batory,FSE2017menzies}.

Traditional approaches to exploring Linux kernel configurations, such as manual tuning, documentation review, and community expertise, are valuable, but inherently limited. These methods struggle with scalability, version-specific knowledge, and incomplete documentation, leaving developers vulnerable to suboptimal decisions and unforeseen interactions between configuration options \cite{acher2022, martin2021a}. Consequently, practitioners lack systematic insights into configuration behavior, hindering efficient kernel optimization~\cite{acher2019b}.

Recent research efforts, including TuxML \cite{acher2019a} and machine learning-based predictive modeling~\cite{acher2019b, martin2021a, alvespereira:hal-02148791, sarkar2015, jamshidi2017, nair2020}, have made strides by sampling configurations and predicting kernel properties like binary size or build success. However, existing datasets typically cover single kernel versions or narrowly defined metrics, limiting their applicability to studies of configuration evolution, cross-version detection of non-functional properties, and complex option interactions~\cite{pereira2020a}.

To address these gaps, we present the \linuxdata, a systematically curated dataset encompassing over 200,000 configurations from multiple Linux kernel versions (including releases from 4.13 to 5.8). We reuse and adapt existing tools, such as TuxML \cite{acher2019a}, to perform large-scale builds using a cluster of machines. Our dataset uniquely captures configuration evolution, build outcomes, and kernel binary sizes, allowing for broader generalization and deeper empirical evidence into the Linux kernel’s configuration space.

To construct the \linuxdata dataset, we rely on extensive random sampling using \texttt{make randconfig}, a standard tool to generate valid Linux kernel configurations while respecting dependency constraints. This approach provides a broad coverage of the configuration space in different versions. Although we do not apply guided sampling, the large number of collected configurations—over 200K—ensures substantial diversity. In future extensions, incorporating guided strategies could complement our current dataset by targeting specific option interactions, performance behaviors, or rarely activated features.

Data-driven methodologies enabled by \linuxdata provide a scalable and systematic approach to understanding kernel configurability. By leveraging extensive historical data and performance measurements, researchers and practitioners can develop predictive models that assist in selecting configurations tailored to specific goals, such as minimizing binary size, optimizing compilation time, or enhancing robustness. Moreover, \linuxdata also supports longitudinal studies that analyze the evolution of configuration impacts over time. This is particularly valuable for system maintainers and developers, where existing documentation may lag behind rapid kernel development, omitting crucial details about new or deprecated options. Thus, the Linux kernel dataset enables significant advancements in predictive modeling, transfer of learning between kernel versions, and automated detection of problematic combinations of options \cite{martin2021a, pereira2020a}. By supporting transfer learning scenarios, the dataset allows models trained on older kernel versions to adapt and make accurate predictions for newer releases, reducing the computational burden of generating training data from scratch. It also supports the investigation of robust sampling strategies and informed feature selection techniques, offering practitioners reliable guidance for kernel configuration decisions in real-world scenarios, such as embedded systems optimization.

The \textit{\linuxdata} dataset\footnote{Available on OpenML \url{https://openml.org/search?type=data&status=active&id=46749}} is publicly available on OpenML \cite{OpenML2013}, along with its associated artifacts, promotes replication, validation and extension of existing research. As we will show and thanks to OpenML, it is possible to import our datasets and explore AI-based techniques with a few lines of Python code.
These empirical data contribute to a structured investigation of how kernel options interact and evolve across versions, offering a valuable resource for researchers studying feature interactions, predictive modeling, and automated configuration tuning. All figures, tables, and code examples presented in this paper are available in the GitHub repository at \url{https://github.com/heraldoborges/tuxkconfig/}. This repository provides a Jupyter notebook that replicates the analyses and visualizations, facilitating reproducibility and further exploration of the \linuxdata dataset.

\section{Dataset Construction}
The construction of this dataset involved a systematic automated process to ensure scalability, consistency, and reproducibility. In the following, we detail the collection and preprocessing phases.



\subsection{Data Collection}
Data were gathered from seven Linux kernel versions: 4.13 (September 2017), 4.15 (January 2018), 4.20 (December 2018), 5.0 (March 2019), 5.4 (November 2019), 5.7 (May 2020), and 5.8 (August 2020). These versions reflect significant evolutionary milestones, including security enhancements (e.g., Meltdown/Spectre mitigations in 4.15) and large-scale updates (e.g., 5.8’s extensive commits) \cite{martin2021a}. We leveraged the TUXML tool \cite{acher2019a} that facilitated the process, automating configuration generation, compilation, and measurement within a Docker environment.

For each version, the randconfig utility generated valid random configurations for the Kconfig dependencies. Compilations aimed at the x86\_64 architecture using GCC 6.3, producing the vmlinux file, a statically linked executable, whose size was measured in megabytes. The number of configurations varied from 92,471 (4.13) to 21,923 (5.8), totaling 243,232 configurations. Compilation times averaged 261 seconds per configuration on 16-core Intel Xeon machines (3.7 GHz, 64 GB RAM), with over 15,000 hours of processing for version 4.13 alone \cite{martin2021a}. Docker ensured portability across heterogeneous clusters.

\subsection{Preprocessing}
Raw data were cleaned and transformed to enhance usability:
\cite{acher2019a}
\begin{itemize}
    \item \textbf{Filtering:} Non-tristate options (300--320 per version, \textit{e.g. integers or strings}) were filtered out as they are marginal in most configuration scenarios and have limited influence on overall kernel behavior \cite{acher2019b, martin2021a}.
    \item \textbf{Encoding:} Tristate values (`y' = enabled, `n' = disabled, `m' = module) were binarized (`y' = 1, `n' or `m' = 0), as `n' and `m' showed similar effects on configuration measurement of non-functional properties \cite{martin2021a}.
    \item \textbf{Derived Feature:} A column summing enabled options (`y') was added, inspired by prior work \cite{acher2019}, aiding predictive modeling.
    \item \textbf{Constant Removal:} Options with uniform values across the dataset were excluded, reducing noise.
    \item \textbf{Baseline Configuration:} The binary sizes of the compiled kernel images (vmlinux) span a wide range of values. For each version, we also include tinyconfig, which produces a minimal kernel to serve as a lower bound in our evaluation.
\end{itemize}

Preprocessing was executed in Python with Pandas and NumPy, producing structured CSV files. The setup ensured hardware independence, leveraging Docker’s standardized environment.

\section{Linux Kernel Dataset}
The dataset comprises CSV files—one file for each version—totaling approximately 2 GB compressed. Each file includes:

\begin{itemize}
    \item \textbf{Version:} Identifier (\textit{e.g.}, "4.13").
    \item \textbf{Date/Time:} Compilation timestamp.
    \item \textbf{Option1, ..., OptionN:} Binary values (0 or 1) for options (\textit{e.g.}, CONFIG\_HAMACHI, CONFIG\_PCI).
    \item \textbf{Binary\_Size (MB):} vmlinux size.
    \item \textbf{Compilation\_Time (s):} Build duration.
    \item \textbf{Sum\_Enabled\_Options:} Count of `y' options.
\end{itemize}

\begin{table}[h]
    \centering
    \caption{Characteristics of Each Version}
    \label{tab:tabversproperties}
    \setcellgapes{2pt}\makegapedcells
    \begin{tabular}{lccc}
        \toprule
        Version & Options & Configurations & \thead{\normalsize Min/Max\\ \normalsize Size (MB)} \\
        \midrule
        4.13 & 12,502 & 92,471 & 7.0 / 1698.1 \\
        4.15 & 9,445 & 39,391 & 11.0 / 1783.8 \\
        4.20 & 10,209 & 23,489 & 11.0 / 2035.5 \\
        5.0  & 10,313 & 19,952 & 11.0 / 1869.5 \\
        5.4  & 10,833 & 25,847 & 11.1 / 1959.4 \\
        5.7  & 11,358 & 20,159 & 11.1 / 1906.0 \\
        5.8  & 11,550 & 21,923 & 11.1 / 1905.1 \\
        \bottomrule
    \end{tabular}
\end{table}

\begin{figure}[h]
    \centering    \includegraphics[width=0.9\columnwidth]{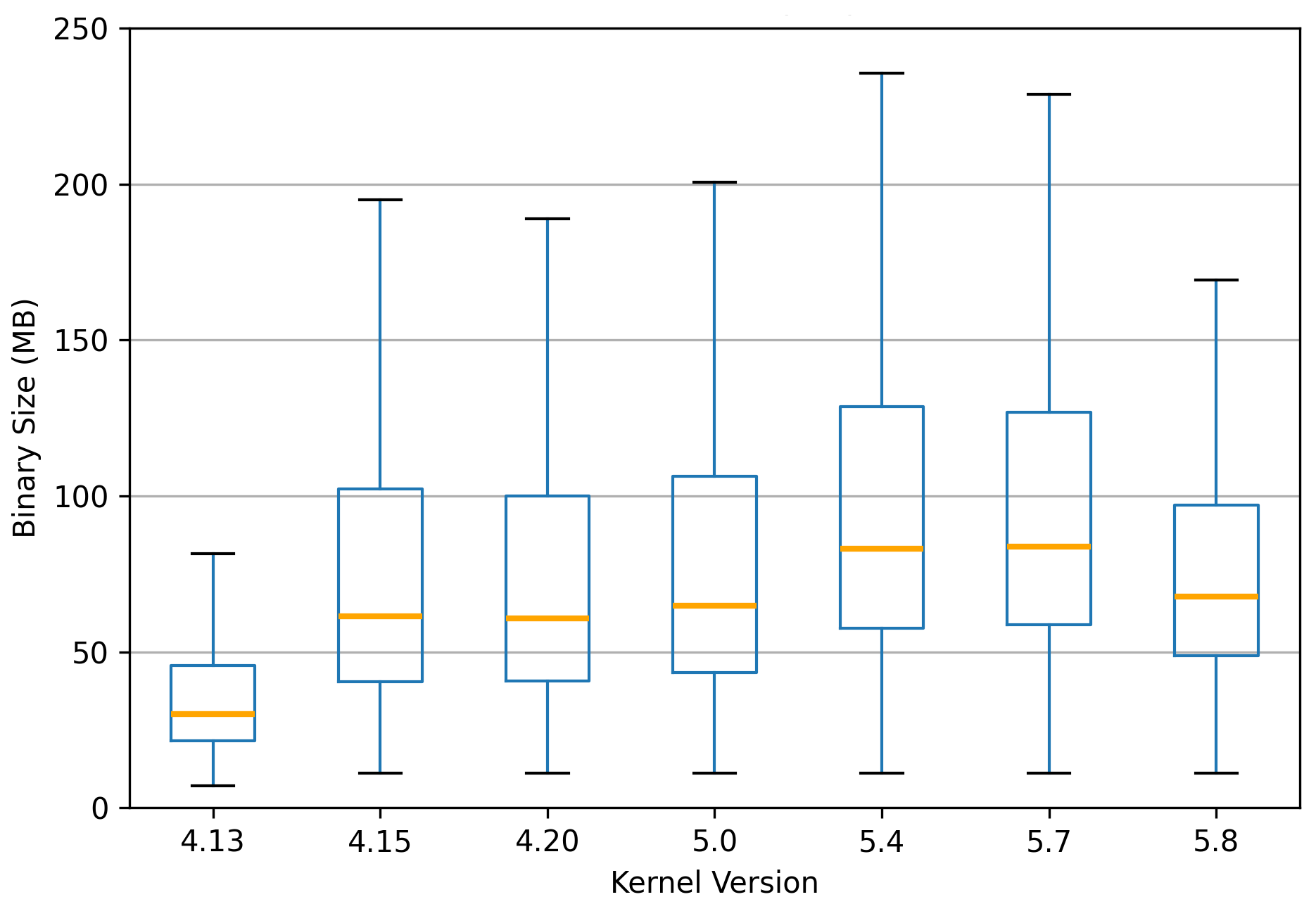}
    \caption{Boxplot of binary size distribution (MB) across versions 4.13 to 5.8.}
    \label{fig:distribution}
\end{figure}

\autoref{tab:tabversproperties} 
and \autoref{fig:distribution} provide a detailed overview of kernel binary size characteristics across versions 4.13 to 5.8. The table summarizes the minimum and maximum kernel binary sizes, highlighting the variations in configurability.  \autoref{fig:distribution} further illustrates the distribution of kernel binary sizes for each version using box plots. The median binary size is represented by the central line within each box, while the interquartile range (IQR) captures the middle 50\% of the data, showing the range where most kernel binary sizes fall. The whiskers extend to indicate the broader distribution, and numerous outliers, represented by densely grouped points beyond the whiskers, highlight configurations with exceptionally large binaries.

Across versions, the distributions reveal a consistent trend. Although most configurations result in moderate-sized kernel binaries, certain configurations lead to significantly larger binaries, often exceeding 800 MB. The presence of numerous outliers suggests that specific kernel options can dramatically increase binary size. This variability highlights the diversity and complexity of kernel configuration effects, reinforcing the importance of dataset-driven analysis, as manual tuning alone may not effectively capture the full spectrum of configuration performance.

The dataset includes multiple measurements of kernel binary size, capturing both the original uncompressed kernel binary size and its compressed forms. Specifically, it records the size of the \texttt{vmlinux} binary—representing the statically linked, uncompressed Linux kernel—along with various compressed kernel image sizes typically used in deployment scenarios. These compressed variants include standard compression algorithms such as gzip (\texttt{GZIP} \texttt{-vmlinux} and \texttt{GZIP} \texttt{-bzImage}), bzip2 (\texttt{BZIP2-} \texttt{vmlinux} and \texttt{BZIP2} \texttt{-bzImage}), LZMA (\texttt{LZMA} \texttt{-vmlinux} and \texttt{LZMA} \texttt{-bzImage}), and XZ (\texttt{XZ-vmlinux}). The \texttt{bzImage} notation, historically referring to a self-extracting compressed kernel image used particularly on x86 architectures, is provided under multiple compression algorithms to facilitate the exploration of kernel sizes under different compression methods. By offering these multiple binary size measurements, the dataset enables researchers and practitioners to analyze and understand how configuration options impact the kernel's footprint, considering different compression choices that are crucial in various practical deployment environments \cite{acher2019b}.

\section{Tasks and Research Opportunities}

The \linuxdata dataset opens a wide range of research opportunities and practical applications due to its comprehensive representation of Linux kernel configurations. Unlike typical tabular datasets—often characterized by limited features and simpler structures—\linuxdata stands out through its scale and complexity, capturing the intricate, interdependent configuration space of an evolving real-world software system. This enables research into both well-explored areas, such as predictive modeling and transfer learning, and emerging opportunities, including hybrid scenarios and tool development, each offering unique insights into kernel configurability and its broader implications \cite{pereira2020b, martin2021a}.

In the following subsections, we present these research directions in detail, highlighting their significance, associated challenges, and potential contributions.
The complete set of scripts for the research directions described, including additional examples for the Optimization and Performance Specialization areas, is available in our supplementary material\footnote{\url{https://github.com/heraldoborges/tuxkconfig/}}, which provides comprehensive support for reproducing and extending the presented techniques.

\subsection{Performance Prediction}
\label{sec:performance-prediction}

Performance prediction in configurable systems aims to estimate non-functional properties, such as execution time, memory usage, or binary size, based on a given configuration. This task is critical in software engineering, where understanding the impact of configuration options enables developers to optimize systems for specific requirements. Research has extensively explored this problem using machine learning (ML) techniques, ranging from regression models to deep learning, to predict results in large configuration spaces \cite{acher2019b,pereira2020a,pereira2020b}. However, the scale and complexity of real-world systems often challenge the precision and efficiency of these methods, particularly when training data is limited or the configuration space is vast and interdependent \cite{acher2019a,martin2021b}. The goal is to map input configurations (e.g., feature selections) to measurable outputs (e.g., performance metrics), typically evaluated using metrics like Mean Absolute Error (MAE) or Mean Relative Error (MRE).

In the context of the Linux kernel, performance prediction becomes particularly compelling due to its highly configurable nature, with over 15,000 options evolving across versions. Previous studies have demonstrated its applicability not only to binary size, but also to compilation time, robustness, or even potential metrics such as boot time \cite{acher2019b,martin2021a}. For example, predicting binary size can guide developers in producing lightweight kernels for resource-constrained environments, while predicting compilation time can optimize build processes \cite{acher2019b}. The challenge lies in the kernel’s combinatorial configuration space and interdependencies, which render exhaustive exploration infeasible and demand ML models that generalize well with minimal training data \cite{acher2019a,acher2022}.

The \linuxdata dataset specifically enables performance prediction by providing over 243,000 configurations across seven Linux kernel versions (4.13 to 5.8), each with detailed binary size measurements. In this context, we focus on predicting the binary size (in MB) of the \texttt{vmlinux} file for a given configuration within a specific version, such as 5.8, without considering cross-version adaptations. The expected input is a vector of binary-encoded options (e.g., \texttt{CONFIG\_HAMACHI = 1} or \texttt{0}), and the output is a continuous value representing the size of the kernel, evaluated using the MAPE to assess the precision of the prediction. To illustrate, we provide a basic Python example using the OpenML dataset of \linuxdata version 5.8 (ID: 46744), employing a Linear Regression baseline:

\begin{lstlisting}
from sklearn.linear_model import LinearRegression
from sklearn.model_selection import train_test_split
import openml
import numpy as np

# Known targets to exclude from training features
targets=["vmlinux","GZIP-bzImage","GZIP-vmlinux","GZIP","BZIP2-bzImage","BZIP2-vmlinux","BZIP2","LZMA-bzImage","LZMA-vmlinux","LZMA","XZ-bzImage","XZ-vmlinux","XZ","LZO-bzImage","LZO-vmlinux","LZO","LZ4-bzImage","LZ4-vmlinux","LZ4"]

# Fetch TuxKConfig v5.8 from OpenML
dataset = openml.datasets.get_dataset(46744)
data,_,_,_=dataset.get_data(dataset_format="dataframe")

# Separate features (options) and target (vmlinux)
X = data.drop(columns=known_targets, errors="ignore")
y = data["vmlinux"]

# Split the data into training and test sets
X_train, X_test, y_train, y_test = train_test_split(X, y, test_size=0.2, random_state=42)

# Train and evaluate a simple linear regression model
model = LinearRegression()
model.fit(X_train, y_train)
predictions = model.predict(X_test)

# Calculate MAPE
mape = np.abs((predictions-y_test)/y_test).mean()*100
print(f"MAPE: {mape:.2f}%")
\end{lstlisting}

In this example, applying to 21,923 configurations from version 5.8, the baseline yields a MAPE of 88.49\%, indicating a significant percentage deviation that suggests room for improving prediction precision with more sophisticated models. This aligns with the findings that simple models struggle with kernel complexity, while more sophisticated approaches (\textit{e.g.} selection of subsets of features or ensemble methods) can improve precision \cite{acher2022,martin2021b}. The scale of the \linuxdata dataset, far exceeding typical tabular datasets in feature count \cite{Varoquaux2021}, poses a unique challenge: Can state-of-the-art predictors scale effectively and maintain accuracy with limited training samples? Researchers can leverage this dataset to benchmark and refine ML techniques, fostering advancements in performance prediction for configurable systems.

\subsection{Transfer Learning}
\label{sec:transfer-learning}

Transfer learning in configurable software systems seeks to harness knowledge acquired from one domain or version to enhance predictions in a related yet distinct context. This approach proves especially valuable when data for a target version is limited, enabling models trained on extensive source data to adapt effectively to new scenarios \cite{martin2021a}. Existing studies underscore its utility in minimizing the need for exhaustive retraining. In addition, these studies highlight improvements in predictive performance in evolving software variants. Nevertheless, the approach faces challenges related to divergent feature sets and shifting data distributions \cite{acher2019b,pereira2020b}. In particular, Martin et al. \cite{martin2021a} demonstrate its promise in navigating expansive configuration spaces, where models must adjust to changing dependencies and options' impacts. The task entails using configuration data and associated performance metrics from a source version as inputs to forecast outcomes—such as size or time—in a target version, with accuracy gauged through metrics like MAPE \cite{acher2019b,martin2021b}.

When applied to the Linux kernel, transfer learning emerges as a strategic method to cope with its dynamic evolution, marked by frequent updates to thousands of configuration options, the introduction of new features, and shifting interdependencies across releases. Previous work has illustrated the potential of leveraging configuration data to predict various system properties, such as binary size and, to a lesser extent, compilation time \cite{martin2021a}. Although more speculative metrics such as boot time remain challenging to model, they represent promising directions for future investigation \cite{martin2021a,acher2019a}. In contrast to the single version focus on performance prediction outlined in Section~\ref{sec:performance-prediction}, this approach capitalizes on historical data to span multiple releases, addressing the difficulty of maintaining predictive accuracy amid kernel evolution \cite{acher2022}. This capability is vital for developers seeking to streamline configuration optimization across successive versions without the burden of collecting fresh data for each release.

The \linuxdata dataset facilitates this task by offering a comprehensive collection of more than 243,000 configurations spanning seven Linux kernel versions (4.13 to 5.8), each accompanied by precise binary size measurements. Here, transfer learning involves training a model on configurations from prior versions, such as 5.4, to predict the \textit{vmlinux} binary size (in MB) for a subsequent version, like 5.8. The input comprises binary-encoded option vectors from the source versions, while the output is the predicted size in the target version, evaluated using MAPE. Below is a Python example utilizing \linuxdata datasets from OpenML—versions 5.4 (ID: 46742) as sources, and 5.8 (ID: 46744) as the target—employing a Hist Gradient Boosting regressor model, known for handling large datasets and capturing complex interactions among options:

\begin{lstlisting}
from sklearn.ensemble import HistGradientBoostingRegressor
from sklearn.model_selection import train_test_split
import openml
import pandas as pd
import numpy as np

# Target columns to exclude from input features
targets=["vmlinux","GZIP-bzImage","GZIP-vmlinux","GZIP","BZIP2-bzImage","BZIP2-vmlinux","BZIP2","LZMA-bzImage","LZMA-vmlinux","LZMA","XZ-bzImage","XZ-vmlinux","XZ","LZO-bzImage","LZO-vmlinux","LZO","LZ4-bzImage","LZ4-vmlinux","LZ4"]

# Load datasets from OpenML
X_504d = openml.datasets.get_dataset(46742)  # v5.4
X_trgd = openml.datasets.get_dataset(46744)  # v5.8
X_504,_,_,_=X_504d.get_data(dataset_format="dataframe")
X_trg,_,_,_=X_trgd.get_data(dataset_format="dataframe")

# Extract targets
y_504 = X_504["vmlinux"]
y_trg = X_trg["vmlinux"]
X_504 = X_504.drop(columns=targets, errors="ignore")
X_trg = X_trg.drop(columns=targets, errors="ignore")

# Feature engineering
for df in [X_504,X_trg]:df["nb_yes"]=df.eq(1).sum(axis=1)

# Align features between source and target
common_feats = X_504.columns.intersection(X_trg.columns)
X_504 = X_504[common_feats]
X_trg = X_trg[common_feats]

# Drop rows with NaN in features or targets
src_df = pd.concat([X_504, y_504], axis=1).dropna()
trg_df = pd.concat([X_trg, y_trg], axis=1).dropna()

X_source = src_df.drop(columns=["vmlinux"])
y_source = src_df["vmlinux"]
X_trg = trg_df.drop(columns=["vmlinux"])
y_trg = trg_df["vmlinux"]

# Train and predict
model = HistGradientBoostingRegressor(random_state=42)
X_source_train, X_source_test, y_source_train, y_source_test = train_test_split(X_source,y_source,test_size=0.2,random_state=42)
model.fit(X_source_train, y_source_train)
predictions = model.predict(X_trg)

# Compute MAPE
mape = np.mean(np.abs((predictions-y_trg)/y_trg))*100
print(f"MAPE (v5.4 - v5.8): {mape:.2f}%")
\end{lstlisting}

This example, drawing on 46,318 configurations from version 5.4 to predict 23,159 configurations in version 5.8, yields a MAPE of approximately 23.31\% \cite{martin2021a}. Although this result remains significantly better than intraversion predictions using simpler linear models (as illustrated in Section\ref{sec:performance-prediction}), it highlights the inherent challenges posed by evolving configuration dependencies and interactions between kernel releases~\cite{martin2021a,acher2022}. These findings reinforce the value of the \linuxdata dataset, with its multiversion scope, as a robust resource to further refine transfer learning methods and deepen our understanding of how kernel configurations evolve across versions. Furthermore, identifying which configuration options most significantly affect prediction accuracy between versions remains crucial, motivating interpretability studies (Section~\ref{sec:interpretability}) to better capture critical configuration elements over time.

\subsection{Interpretability}
\label{sec:interpretability}

Interpretability in machine learning models applied to configurable systems seeks to uncover which configuration options most significantly influence system behavior, offering insights into their relative importance. This task is essential for demystifying complex predictive models, where opaque algorithms—such as neural networks or ensemble models like random forests—often obscure the contribution of individual features \cite{pereira2020b,martin2021b}. Prior research has grappled with the instability of feature importance measures, noting that results can fluctuate across model runs due to randomness or interdependencies among options \cite{acher2019a,acher2019b}. Acher et al. \cite{acher2022} emphasize that identifying a small subset of pivotal features can explain much of the variability in system outcomes, yet achieving consistent and actionable insights remains elusive in vast configuration spaces \cite{acher2022}.

In highly configurable systems like the Linux kernel, interpretability shifts focus to pinpointing which of the over 15,000 options drive key properties—be it binary size, compilation time, or other performance indicators. Beyond merely predicting outcomes, as explored in Sections~\ref{sec:performance-prediction} and~\ref{sec:transfer-learning}, this task aims to illuminate why certain predictions hold, helping developers prioritize configuration adjustments \cite{martin2021a}. For instance, understanding whether \texttt{CONFIG\_PCI} outweighs \texttt{CONFIG\_HAMACHI} in affecting kernel size can guide optimization efforts. However, the scale of the kernel and the intricate dependencies of the options complicate this analysis, as standard methods, such as the importance of the features of tree-based models, may yield inconsistent rankings in executions \cite{martin2021b}. This challenge, raised in previous discussions, underscores the need for robust techniques to extract and filter meaningful patterns from such expansive datasets.

Using version 5.8 of the \linuxdata dataset, available in OpenML (ID: 46744) and containing detailed binary size data per configuration, we focus on identifying the options that most influence the size of the generated \texttt{vmlinux}. Interpretability here entails taking binary-encoded option vectors as input and producing a weighted ranking of influential configuration options as output. To illustrate this approach, we apply a Random Forest regressor and assess the configuration options via their feature importance scores:

\begin{lstlisting}
from sklearn.ensemble import RandomForestRegressor
from sklearn.model_selection import train_test_split
import openml
import pandas as pd
import numpy as np

targets=["vmlinux","GZIP-bzImage","GZIP-vmlinux","GZIP","BZIP2-bzImage","BZIP2-vmlinux","BZIP2","LZMA-bzImage","LZMA-vmlinux","LZMA","XZ-bzImage","XZ-vmlinux","XZ","LZO-bzImage","LZO-vmlinux","LZO","LZ4-bzImage","LZ4-vmlinux","LZ4"]

# TuxKConfig v5.8 (dataset ID: 46744)
dataset = openml.datasets.get_dataset(46744)
X,_,_,_=dataset.get_data(dataset_format="dataframe")
y = X["vmlinux"]

# Remove known target
X = X.drop(columns=targets, errors="ignore")


# Train model for feature importances
X_train, X_test, y_train, y_test = train_test_split(X, y, test_size=0.2, random_state=42)
model=RandomForestRegressor(100,random_state=42)
model.fit(X_train, y_train)
y_pred = model.predict(X_test)

# MAPE
mape = np.mean(np.abs((y_test-y_pred)/y_test))*100
print(f"MAPE: {mape:.2f}%")

# Compute feature importances from the trained model
importances = model.feature_importances_
features = X.columns

# The top 5 most influential configuration options
top_indices = importances.argsort()[-5:][::-1]
top5 =[(features[i],importances[i]) for i in top_indices]

for feature,score in top5:print(f"{feature}:{score:.4f}")
\end{lstlisting}

Running this analysis on 21,923 configurations from version 5.8 reveals that options such as \texttt{DEBUG\_INFO} (0.2067), \texttt{active\_options} (0.2004), and \texttt{DEBUG\_INFO\_REDUCED} (0.1647) strongly influence kernel binary size. Additionally, \texttt{DEBUG\_INFO\_}\texttt{SPLIT} (0.1235) and \texttt{X86\_} \texttt{NEED\_RELOCS} (0.0829) also significantly contribute to variability. These findings align with prior observations that a small subset of configuration options tends to dominate kernel size variability \cite{acher2022}. The cumulative importance of these top-five features accounts for approximately 78\% of the total variance, clearly highlighting their critical role in determining kernel size. The predictive model achieved an MAPE of approximately 11.04\%, demonstrating a satisfactory predictive accuracy. However, the consistency of feature importance across different kernel versions remains a recognized challenge \cite{martin2021b}. Building upon insights from transfer learning discussed in Section~\ref{sec:transfer-learning}, which illustrated evolving option influences between kernel releases, this interpretability analysis provides a foundational step toward feature selection. By pinpointing these influential options, we set the stage for efficiently reducing the configuration space, a task explored further in Section~\ref{sec:feature-selection} through targeted feature selection methods.

\subsection{Feature Selection}
\label{sec:feature-selection}

Feature selection addresses the challenge of paring down the configuration options in complex software systems to a subset that retains essential predictive power while easing computational demands. This task emerges as a cornerstone in managing high-dimensional spaces, where analyzing all possible features exhaustively becomes computationally prohibitive \cite{pereira2020b}. Research has shown that many options contribute redundantly or minimally to system behavior, and by isolating a core set, one can streamline analysis without sacrificing accuracy \cite{acher2019b,acher2022}. Acher et al. \cite{acher2022} illustrate this by demonstrating that a fraction of features can account for most performance variation, though selecting such a subset requires balancing relevance against the risk of overlooking subtle interactions \cite{martin2021b}. The process begins with a complete configuration dataset as input and yields a reduced feature set as output, typically assessed by the prediction accuracy achieved with the selected options.

Within the realm of configurable systems like the Linux kernel, feature selection tackles the daunting scale of over 15,000 options, aiming to simplify analyses of properties such as binary size, compilation time, or system robustness. Although interpretability (Section~\ref{sec:interpretability}) highlights which options wield the greatest influence, feature selection shifts the focus to curating a practical subset for downstream tasks like prediction or tuning \cite{acher2019b}. This reduction proves critical when adapting models across versions or crafting efficient kernels, yet the kernel’s dense web of dependencies complicates the endeavor—discarding features may inadvertently disrupt key relationships \cite{martin2021a}. The challenge lies in devising methods that scale effectively and remain stable in such complex scenarios.

Using version 5.8 of the \linuxdata dataset from OpenML (ID: 46744), feature selection targets the reduction of options for predicting \texttt{vmlinux} binary size. The input consists of binary-encoded configuration vectors from 21,923 instances, and the output is a trimmed set of influential options, evaluated using MAPE of a model trained on this reduced subset. Below is a Python example employing a Random Forest-based for both feature selection and prediction to identify critical configuration options and assess their predictive performance.

\begin{lstlisting}
from sklearn.ensemble import RandomForestRegressor
from sklearn.feature_selection import SelectFromModel
from sklearn.model_selection import train_test_split
import openml
import pandas as pd
import numpy as np

targets=["vmlinux","GZIP-bzImage","GZIP-vmlinux","GZIP","BZIP2-bzImage","BZIP2-vmlinux","BZIP2","LZMA-bzImage","LZMA-vmlinux","LZMA","XZ-bzImage","XZ-vmlinux","XZ","LZO-bzImage","LZO-vmlinux","LZO","LZ4-bzImage","LZ4-vmlinux","LZ4"]

dataset = openml.datasets.get_dataset(46744)
X, _, _, _ = dataset.get_data(dataset_format="dataframe")
y = X["vmlinux"]
X = X.drop(columns=targets, errors="ignore")

# Split data
X_train, X_test, y_train, y_test = train_test_split(X, y, test_size=0.2, random_state=42)

# Use RandomForest for feature selection
selector = SelectFromModel(RandomForestRegressor(n_estimators=100, random_state=42))
selector.fit(X_train, y_train)
selected_features = X.columns[selector.get_support()]

# Evaluate with selected features
model = RandomForestRegressor(n_estimators=100, random_state=42)
model.fit(X_train[selected_features], y_train)
y_pred = model.predict(X_test[selected_features])

# MAPE evaluation
mape = np.mean(np.abs((y_test-y_pred)/(y_test)))*100
print(f"MAPE - Feature selection: {mape:.2f}%")
\end{lstlisting}

This example effectively reduced the initial 11,531 options to approximately 130
by employing the Select From Model feature selection technique in combination with a Random Forest Regressor. This method automatically selects features whose importance, calculated by the Random Forest model, exceeds a predefined threshold. Specifically, we adopted the default threshold ("mean"), thus retaining only options whose importance exceeded the average value. These results align with findings by Acher et al. \cite{acher2022}, who similarly observed that a relatively small subset of options is often sufficient to achieve accurate kernel size predictions. A subsequent model trained on this reduced subset achieved a Mean Absolute Percentage Error (MAPE) of approximately 9.74\%, demonstrating an improvement in predictive accuracy compared to the 11.04\% obtained using the complete set of features presented in Section~\ref{sec:interpretability}. This suggests that the feature selection process successfully identified and retained the most informative options, resulting in both a simplified and a more accurate predictive model. Although cross-version consistency requires further investigation \cite{martin2021b}, this substantial feature reduction, based on the interpretability analysis of Section~\ref{sec:interpretability}, enables optimizing minimal kernels by prioritizing critical options such as \texttt{SSB\_DRIVER\_PCICORE}, \texttt{BNX2}, \texttt{DMA\_COHERENT\_POOL}, \texttt{ARCH} \texttt{\_HAS\_FORCE\_DMA\_UNENCRYPTED}, and \texttt{PM\_DEVFREQ}, as illustrated in size-constrained scenarios discussed in Section~\ref{sec:optimization}, where the objective changes to achieving compact, yet effective kernel configurations.

\subsection{Optimization}
\label{sec:optimization}

Optimization in software systems entails identifying configurations that best satisfy predefined objectives, such as minimizing resource usage or maximizing efficiency. This endeavor holds practical significance across engineering domains, where the need to balance performance constraints against system requirements drives innovation \cite{pereira2020a}. Literature has framed it as a search problem within expansive configuration landscapes, where exhaustive enumeration falters due to combinatorial growth \cite{acher2019b,martin2021b}. Acher et al. \cite{acher2019b} suggest that predictive models can inform this process by highlighting factors tied to specific outcomes, although success depends on navigating trade-offs and interdependencies effectively \cite{acher2022}. The task takes a configuration dataset as input and aims to deliver an optimal configuration as output, often measured by the degree to which it meets the target criterion, such as a minimal value for a given metric.

In the domain of configurable systems, optimization focuses on tailoring software to meet precise goals, considering the multitude of available options and their intricate relationships. For the Linux kernel, with its thousands of configuration choices, this translates to crafting setups that excel in diverse aspects—ranging from binary size and compilation speed to operational robustness \cite{martin2021a}. Unlike earlier tasks that predict or interpret outcomes (Sections~\ref{sec:performance-prediction}--\ref{sec:interpretability}), optimization seeks actionable solutions, leveraging insights into influential factors to guide the search process \cite{acher2019b}. The complexity arises from the sheer scale of possibilities, where dependencies among options can obscure straightforward paths to an ideal configuration.

Within the \linuxdata dataset, covering seven kernel versions, optimization narrows to minimizing the \texttt{vmlinux} binary size—a critical goal for resource-constrained environments. Here, the input comprises detailed configuration records with associated size measurements, and the output is a single configuration that produces the smallest size, assessed directly by its binary size in MB (\textit{e.g.}, approaching the 7.3 MB of \texttt{tinyconfig}, a known minimal benchmark)~\cite{randrianaina2023}. Building on the reduced feature sets identified in Section~\ref{sec:feature-selection}, this task could employ techniques such as genetic algorithms or gradient-based searches to pinpoint such a configuration, as suggested by the emphasis \cite{acher2019b} on size-influencing options \cite{acher2019b,acher2022}. While practical implementations might reveal a kernel under 10 MB, the dataset’s extensive coverage supports exploring trade-offs—say, between size and compilation time—laying a foundation for tailoring kernels to specific performance ranges, as pursued in Section~\ref{sec:performance-prediction}.

\subsection{Performance Specialization}
\label{sec:performance-specilization}

Performance specialization in software systems entails identifying configurations that satisfy specific performance thresholds, such as a predefined range of resource use or operational efficiency. This task is crucial in those situations where systems must meet exacting constraints beyond general optimization \cite{martin2021b}. Research positions it as a constrained search within intricate configuration domains, employing methods such as restricted exploration or reinforcement learning to balance diverse factors against set targets \cite{acher2019b,martin2021b}. Martin et al. \cite{martin2021b} affirm its practicality, showing that such approaches can isolate configurations within the desired bounds, although complexity persists in expansive spaces. The process uses a configuration dataset as input to yield configurations that meet the target range, evaluated by their adherence to the specified limits.

In configurable systems such as the Linux kernel, with its numerous configuration possibilities, performance specialization targets set-up within defined intervals, covering binary size, compilation efficiency, or runtime stability \cite{martin2021a}. Unlike optimization’s minimization goal (Section~\ref{sec:optimization}), it seeks a spectrum, such as a 20 to 50 MB kernel size for an IoT device \cite{martin2021b}. The dependencies between options can push configurations beyond the desired range, posing a key challenge \cite{martin2021b}.

The \linuxdata dataset supports this task by providing configuration records and size data to identify \texttt{vmlinux} sizes within a predefined range—\textit{e.g.}, 20 to 50 MB. The input comprises configuration details, and the output is a subset that fits the interval, evaluated by binary size in MB. Extending Section~\ref{sec:optimization}’s minimal-size focus, techniques like constrained search could pinpoint such configurations \cite{martin2021b}, setting the stage for integrating multiple objectives as described in Section~\ref{sec:hybrid-scenarios}.

\subsection{Hybrid Scenarios}
\label{sec:hybrid-scenarios}

Hybrid scenarios in configurable systems involve combining different tasks or techniques to address complex and nuanced challenges more effectively. Instead of treating tasks such as performance prediction (Section~\ref{sec:performance-prediction}), transfer learning (Section~\ref{sec:transfer-learning}), interpretability (Section~\ref{sec:interpretability}), feature selection (Section~\ref{sec:feature-selection}), or optimization (Section~\ref{sec:optimization}) in isolation, researchers can integrate them to achieve more cost-effective outcomes. For instance, feature selection (Section~\ref{sec:feature-selection}) can be paired with interpretability (Section~\ref{sec:interpretability}) to focus on the most influential configuration options while reducing dimensionality, with the hope of improving accuracy in performance prediction (Section~\ref{sec:performance-prediction}). Similarly, combining interpretability or feature reduction with transfer learning (Section~\ref{sec:transfer-learning}) can potentially enhance inter-version adaptability by selecting stable, meaningful features. In optimization tasks (Section~\ref{sec:optimization}), leveraging interpretability insights can guide search strategies. 

The Linux kernel, with its vast and evolving configuration space, is an interesting subject for such hybrid approaches since building and measuring configurations is very costly. For example, one might reduce features in version 4.13 using interpretability techniques, then transfer this knowledge to version 5.8—though this raises challenges such as preserving critical dependencies across versions \cite{martin2021b}. Extending the constrained focus of performance specialization (Section~\ref{sec:performance-specilization}), the \linuxdata dataset enables testing these integrated techniques at scale \cite{acher2019b}, laying the groundwork for the development of practical tools in Section~\ref{sec:building-tools}.

\subsection{Building Tools out of the Dataset}
\label{sec:building-tools}

A variety of practical tooling for developers, researchers, and maintainers working with the Linux kernel can be envisioned on top of \linuxdata. For instance, the dataset can support the development of improved documentation and exploration tools for Kconfig options. Currently, documentation is often sparse and rarely details the impact of configuration options -- especially in terms of binary size. When such information is provided, it is typically approximate and not consistently updated across kernel versions~\cite{ passos2018, acher2022}. Tools that visualize the effect of options on binary size, identify frequently co-occurring options, or track how options evolve across versions would be highly valuable. These capabilities could help both newcomers and experienced developers better navigate the vast configuration space, revealing hidden dependencies and deprecated features.

In addition, \linuxdata can enable configuration automation through recommendation systems and optimization assistants. A recommender trained on historical configurations could suggest sensible defaults based on common usage patterns or previously successful builds. Similarly, tuners could leverage the dataset to propose optimized configurations tailored for size, performance, or other constraints. Finally, predictive models trained on the dataset -- for example, to estimate binary size -- could support more informed configuration workflows and fully taking advantage of the Linux kernel’s configurability.

\subsection{Interests for the Machine Learning Community}

Tabular data remains a critical frontier for ML, where regression and classification tasks test resilience against real-world complexity \cite{grinsztajn2022tree}. Research underscores the persistent edge of tree-based models over deep learning in such contexts, driven by data sets with high count of features and intricate patterns \cite{acher2019b,pereira2020b}. Varoquaux et al. \cite{grinsztajn2022tree} argue that tabular datasets, unlike simples benchmarks like Iris with its four features, challenge ML methods with scale and heterogeneity, pushing the boundaries of generalization and efficiency \cite{grinsztajn2022tree,martin2021b}. This intersection of data characteristics and algorithmic demands offers fertile ground for advancing ML methods.

In configurable systems, these challenges amplify as numerous options and their dependencies shape performance outcomes. For the Linux kernel, with its hundreds of configuration choices, the \linuxdata dataset provides a tabular resource spanning 243,232 instances in seven versions, capturing binary size and compilation times for \texttt{x86\_64} \cite{acher2019b}. With a feature dimensionality far exceeding that of typical tabular datasets, it probes the capacity of methods like gradient boosting or deep learning to address high-dimensional regression, aligning with concerns raised in Varoquaux et al. \cite{grinsztajn2022tree}'s analysis of tabular data limitations. Beyond software engineering tasks (Sections~\ref{sec:performance-prediction}--\ref{sec:hybrid-scenarios}), it enables ML researchers to probe algorithmic scalability and robustness, bridging domain-specific analysis with broader methodological innovation for practical tool development (Section~\ref{sec:building-tools}).

\section{Related Work}

In addition to Linux, several datasets have been established for performance prediction and optimization in configurable systems~\cite{alvespereira:hal-02148791,jamshidi2018learning,DBLP:conf/sigsoft/SiegmundGAK15,quinton2020evolution,beetle, DBLP:conf/wosp/ValovPGFC17, nair2017, FSE2017batory,FSE2017menzies, bessa2025}. Systems like Apache HTTP Server~\cite{sarkar2015}, x264~\cite{nair2020}, SQLite~\cite{jamshidi2017} and Berkeley DB~\cite{sarkar2015} have been extensively used to evaluate ML techniques across diverse configuration spaces. These benchmarks often provide sampled measurements of performance metrics (e.g., execution time, throughput) across hundreds of configurations. Similarly, systems such as SPEAR~\cite{jamshidi2017}, SaC~\cite{jamshidi2017}, and ExaStencils~\cite{peng2021veer} support high-dimensional exploration of configuration landscapes and serve as testbeds for deep learning or surrogate-assisted optimization methods.

In addition to performance-focused datasets, feature model repositories such as SPLOT~\cite{pereira2014} and the recent UVL corpus~\cite{sundermann2024uvl} offer structural insights into variability in domains such as automotive, embedded systems and IoT. These collections are widely used for benchmarking variability analysis (\textit{e.g.}, with satisfiability solvers). Moreover, configurable systems like MySQL~\cite{huang2016skoll,eddington2018skoll} have been used to build datasets of failure-inducing configurations, supporting fault localization and test prioritization tasks. Abal et al.~\cite{Abal2018VariabilityBugs,Abal2014LinuxVariabilityBugs} introduce the Variability Bugs Database (VBDb), a benchmark designed to document and analyze real-world variability bugs—such as feature interaction issues—in highly configurable systems, with a focus on the Linux kernel and other large-scale software.

Numerous works study Linux and its huge configuration space~\cite{gazilloFSE, chico2025, gazillo2021, DBLP:conf/splc/Fernandez-Amoros24, DBLP:conf/icse/FranzBFNG21, 10.1145/1966445.1966451, DBLP:conf/splc/MortaraC21}.
Previous efforts on Linux kernel data include the compilation of over 95,000 configurations for version 4.13 by Acher et al.~\cite{acher2019}, which focused on predicting binary size, and Melo et al.'s collection of 21,000 configurations~\cite{melo2016} for analyzing compilation warnings in a single version. In contrast, our dataset spans multiple kernel versions (4.13 to 5.8), capturing the evolutionary dynamics of configuration options over time. Studies like Passos et al.~\cite{passos2018} investigated feature model evolution but lacked performance metrics, while Sharifdeen et al.~\cite{shariffdeen2024} explored software reuse without a specific focus on Linux. Broader ML research, such as that of Martin et al.~\cite{martin2021a}, has utilized similar datasets for transfer learning applications. However, our work uniquely addresses heterogeneous variability across both space and time (version changes) dimensions. Existing works mainly consider transfer learning with the same feature set (\textit{e.g.}, ~\cite{jamshidi2017, alvespereira:hal-02148791}.
Beyond transfer learning, several use-cases over our dataset can be considered (\textit{e.g.}, feature selection, interpretability) to assess or devise AI-based techniques.


To the best of our knowledge, \linuxdata is the largest existing curated dataset of configurable systems, not limited to the Linux kernel. Spanning seven kernel versions (4.13 to 5.8), involving 10K+ features, and covering over 200K+ configurations, it exceeds the scale of prior datasets that typically focus on a single version or system. As we have described, our dataset calls to investigating whether AI-based techniques can effectively scale to address key challenges in software configurable systems, including performance prediction, optimization, specialization, transfer learning, and interpretability.

\section{Conclusion}
In this paper, we introduced \linuxdata a large dataset of 243,861 Linux kernel configurations across 7 versions from 4.13 to 5.8, spanning three years of evolution. The dataset uniquely tracks the evolving configuration space with build outcomes and binary sizes. As described, it is suitable for research in various tasks (\textit{e.g.}, performance prediction, transfer learning, identification of influential configuration options) that typically involve the use of AI techniques (\textit{e.g.}, supervised machine learning, sampling strategies). The overall challenge is to find cost-effective solutions capable of accurately predicting and inferring insights out of the huge configuration space of the Linux kernel.
We have made an effort to make the dataset available in OpenML and showcase how to use it in a few lines of Python.
Beyond the software engineering community, we believe \linuxdata is of interest for researchers interested in tabular data -- the scale of our dataset as well as the evolution of the feature set is unique and thus challenging.

Our future work plan is to expand \linuxdata further with more versions and on other architectures, hopefully delivering a growing dataset over time for the Linux research community.
We also plan to reproduce existing works in the software engineering and AI community using our dataset, leading to a leaderboard and actionable picture of the effectiveness of state-of-the-art solutions with regard to various tasks.


\section*{Acknowledgements}
This research was partially funded by the Brazilian funding agencies CAPES (Grant 88881.879016/2023-01), FAPESP (Grant 2023/00811-0) and the Binational Cooperation Program CAPES/COFECUB (Ma1036/24). We also acknowledge the Brazilian company Stone for the financial support.



\end{document}